\documentclass[aps,groupaddress,superscriptaddress,reprint,amsmath,amssymb,graphicx]{revtex4-1}
\usepackage{graphicx}
\usepackage{physics}
\usepackage{bm}
\usepackage{bbold}
\usepackage{color}
\begin{document}
\preprint{APS/123-QED}

\title{Tuning transduction from hidden observables to optimize information harvesting}

\author{Giorgio Nicoletti}
\affiliation{ECHO Laboratory, École Polytechnique Fédérale de Lausanne, Lausanne, Switzerland}
\author{Daniel Maria Busiello}
\affiliation{Max Planck Institute for the Physics of Complex Systems, Dresden, Germany}

\begin{abstract}
\noindent Biological and living organisms sense and process information from their surroundings, typically having access only to a subset of external observables for a limited amount of time. In this work, we uncover how biological systems can exploit these accessible degrees of freedom (DOFs) to transduce information from the inaccessible ones with a limited energy budget. We find that optimal transduction strategies may boost information harvesting over the ideal case in which all DOFs are known, even when only finite-time trajectories are observed, at the price of higher dissipation. We apply our results to red blood cells, inferring the implemented transduction strategy from membrane flickering data and shedding light on the connection between mechanical stress and transduction efficiency. Our framework offers novel insights into the adaptive strategies of biological systems under non-equilibrium conditions.
\end{abstract}

\maketitle

\noindent Understanding how biological systems sense and process information from their surroundings is a long-standing question \cite{bialek2005physical,tkavcik2016information,kussell2005phenotypic,bauer2023information}. A fundamental problem is that these systems cannot typically access all the degrees of freedom (DOFs) characterizing the external world. Rather, they have to rely on a subset of stochastic observables transmitting such external information to their internal processes. This transduction mechanism is made even more challenging by the intrinsic spatial and temporal limitations of the accessible trajectories - e.g., chemical concentrations \cite{kaizu2014berg, mora2019physical, berg1977physics} or positions in space \cite{evans1986thermal,nicoletti2021mutual,nicoletti2022mutual, nicoletti2022information}. How transduction allows for efficient information harvesting is far from being understood. Furthermore, several works that investigated theoretical and experimental signaling pathways highlighted that any information-processing operation inevitably requires energy \cite{cheong_tnf,sagawa2009minimal,lan2012energy,govern2014energy}. In addition, recent advances uncovered theoretical limits on sensing of noisy variables \cite{kaizu2014berg,mora2019physical,malaguti2021theory,bauer2023information}. However, very little is known about how much information on inaccessible observables is contained in the accessible ones. This is especially relevant since all biological systems operate with a limited energy budget. As a consequence, they have to tune their transduction strategies accordingly to achieve the maximal amount of information on the inaccessible or hidden DOFs. Moreover, the presence of multiple timescales \cite{nicoletti2023propagation, nicoletti2023information, bo2017multiple, mariani2022disentangling}, the non-reciprocity of the interactions with the environment \cite{klapp2023non, busiello2023unraveling}, and the intrinsic activity of biological systems \cite{fang2019nonequilibrium} make the identification of optimal strategies a formidably complex task.

In this Letter, we tackle these problems and show that optimal transduction strategies may boost information harvesting over the ideal case in which all DOFs are known. We start by studying a hierarchical model, where a particle is coupled non-reciprocally to an intermediate observable that relays the information of a hidden DOF. The transduction strategy is implemented by tuning the coupling strength. We identify the regimes in which the dissipation due to the unobservable DOF either prevents transduction or makes it inefficient over the ideal case. Further, we show that, on average, efficient transduction inevitably causes an increase in both information variance and dissipation rate. However, biological systems usually extract information from stochastic trajectories on the fly. Hence, we explore how the measurement time affects the optimal transduction coupling through a simple adaptive dynamics. Finally, we extend our framework to analyze experimental data on red blood cells (RBCs). We quantify how much information on cytoskeleton activity is transduced into membrane flickering, unraveling the connection between transduction strategies and dissipation and highlighting intriguing differences in performance depending on mechanical conditions.

\begin{figure}[b]
    \centering
    \includegraphics[width=\columnwidth]{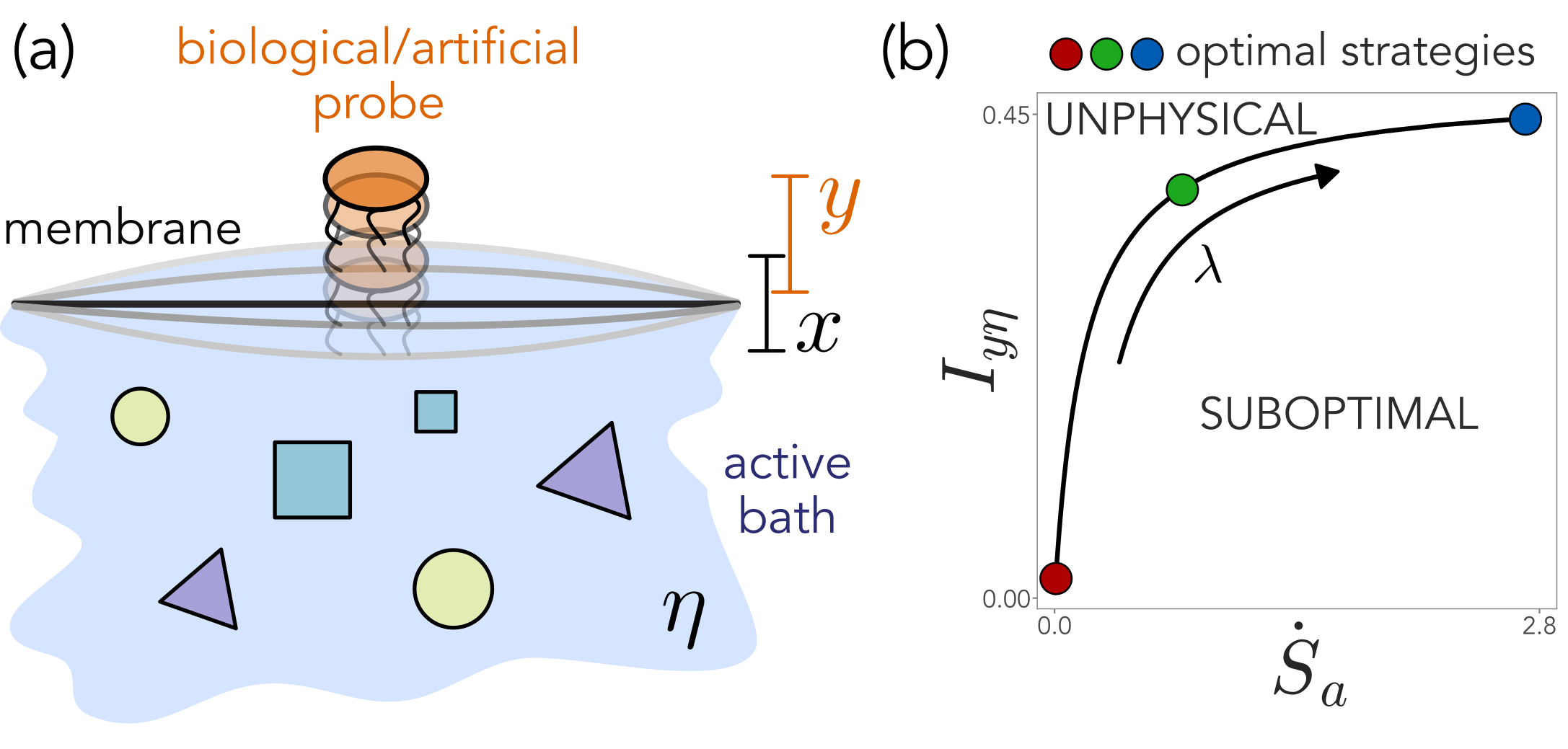}
    \caption{(a) Sketch of the model. Membrane fluctuations, $x$, transduce information from an active bath, $\eta$, to a probe, $y$. (b) Pareto front. Mutual information between $y$ and $\eta$ is maximized while minimizing dissipation. Transduction strategies correspond to different points on the optimal front. In this figure, $\theta_\eta = 5$, $\sigma = 1$.}
    \label{fig:1}
\end{figure}

\begin{figure*}[th!]
    \centering
    \includegraphics[width=\textwidth]{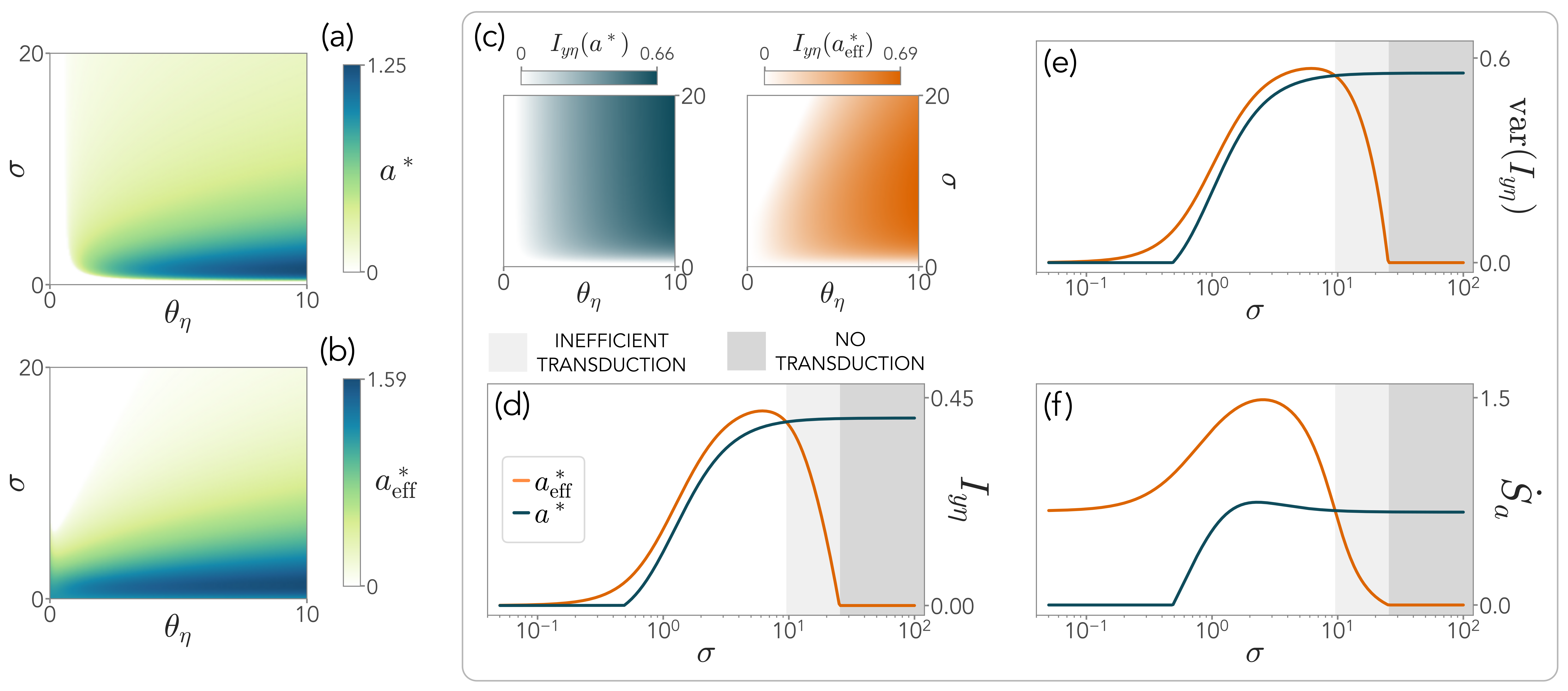}
    \caption{(a-b) Optimal coupling in the ideal case ($a^*$) and considering only the accessible DOFs ($a^*_{\rm eff}$) as a function of membrane-bath interaction, $\sigma$, and activity timescale, $\theta_\eta$, for $\lambda = 0.9$. (c) Target information $I_{y\eta}(a^*)$ (blue) and transduced information $I_{y\eta}(a^*_{\rm eff})$ (orange), for $\lambda = 0.9$. (d-f) Optimal transduction can boost information harvesting over the ideal case (d), at the expense of precision (e) and dissipation (f). Regimes of inefficient transduction appear at increasing $\sigma$. At a critical membrane-bath coupling, transduction is ineffective, thus no information can be harvested and dissipation comes only from the active bath. Here, $\lambda = 0.9$, $\theta_\eta = 5$.}
    \label{fig:2}
\end{figure*}

To fix the ideas, consider a membrane flickering due to the activity of an internal bath, $\eta$ (e.g., the cytoskeleton for RBCs \cite{turlier2016equilibrium}). The orthogonal membrane undulations, $x$, can be measured by an external biological or artificial probe, $y$, through a coupling parameter, $a$. This coupling may very well be non-reciprocal, as the probe may not influence membrane motion. Thus, we have the following hierarchical active model (Fig.~\ref{fig:1}a):
\begin{eqnarray}
    \label{model}
    \tau_y \dot{y} &=& - y + a x + \sqrt{2 \tau_y D_y} \xi_y(t) \nonumber \\
    \tau_x \dot{x} &=& - x + \sigma \eta + \sqrt{2 \tau_x D_x} \xi_x(t) \\
    \tau_\eta \dot{\eta} &=& - \eta + \sqrt{2 \tau_\eta D_\eta} \xi_\eta(t) \;, \nonumber
\end{eqnarray}
where $\sigma$ is the bath-membrane coupling, $\xi_i$ a zero-mean Gaussian white noise, $D_i$ the diffusion coefficient, and $\tau_i$ the typical timescale, with $i = x, y, \eta$. Since the probe has to harvest information on $\eta$ without being able to interact with it directly, we focus on the role of the indirect coupling $a$. That is, the biological probe must tune its coupling to the membrane to optimize the transduction of information from the bath. Without loss of generality, we fix $D_x = D_y = D_\eta = 1$ and $\tau_x = \tau_y = \tau = 1$. A leading role is played by both $\sigma$, which controls the membrane-bath dissipation, and $\tau_\eta \equiv \tau \theta_\eta$, which dictates on which timescale activity is not averaged out \cite{nicoletti2023propagation}.

The task of the probe is to extract information on the bath. Consider first the ideal case in which all DOFs are known and accessible. In this scenario, the probe particle can directly maximize the mutual information between $\eta$ and itself, $I_{y\eta}$, and minimize the dissipation induced by its coupling to the membrane, $\dot{S}_a = \dot{S}_{\rm tot} - \dot{S}_{\rm tot}|_{a = 0}$. This amounts to finding the coupling $a^*$ that maximizes the Pareto functional $\mathcal{L}(a)$ \cite{seoane2015phase}:
\begin{equation}
    a^* = \arg\max_a \underbrace{ \left( \lambda I_{y\eta} - (1-\lambda) T \dot{S}_a \right)}_{\mathcal{L}(a)} \;,
    \label{eqn:functional}
\end{equation}
where $T$ is taken as unit time and $0 < \lambda < 1$. Since information and dissipation are usually in trade-off \cite{lan2012energy}, an optimal front characterizing transduction in different regimes naturally emerges (Fig.~\ref{fig:1}b).

The parameter $\lambda$ in Eq.~\eqref{eqn:functional} represents the strategy that the probe may implement. For very small $\lambda$, the probe is acting to preferentially minimize dissipation, while a very high $\lambda$ denotes an information-driven strategy. 
Assuming that the system operates at stationarity, we can write the mutual information between $y$ and $\eta$ as
\begin{equation}
    I_{y\eta} = \frac{1}{2} \log\frac{\Sigma_{yy} \Sigma_{\eta\eta}}{\det(\mathbf{\Sigma}_{y\eta})}
    \label{info}
\end{equation}
where $\mathbf\Sigma$ is the covariance matrix of the system, and $\mathbf{\Sigma}_{ij}$ its corresponding submatrix on variables $i$ and $j$. Similarly, the dissipation can be written as:
\begin{equation}
    \dot{S}_{\rm tot} = \Tr\left(\mathbf{D}^{-1} \mathbf{A} \mathbf{\Sigma} \mathbf{A}^T\right) - \Tr\left(\mathbf{A}\right)
    \label{sdot}
\end{equation}
where $\mathbf{A}$ and $\mathbf{D}$ are, respectively, the interaction and the diffusion matrices associated with Eq.~\eqref{model}. We present an explicit derivation of these expressions in the Supplemental Material \cite{supplemental_material}.

However, this ideal scenario cannot be realized in practice. Being only coupled to $x$, the probe does not have access to the evolution of $\eta$. As such, both $I_{y\eta}$ and $\dot{S}_a$ cannot be estimated directly, nor the functional in Eq.~\eqref{eqn:functional}. From the stochastic trajectories seen by the probe, the only measurable quantities are instead its information shared with the membrane, $I_{xy}$, and the dissipation stemming from the observable DOFs, $\dot{S}_{xy}$. Therefore, the probe must optimize an effective functional, $\mathcal{L}_{\rm eff}$, to determine its effective transduction coupling:
\begin{equation}
    a^*_{\rm eff} = \arg\max_a \underbrace{ \left( \lambda I_{xy} - (1-\lambda) T \dot{S}_{xy} \right)}_{\mathcal{L}_{\rm eff}(a)} \;.
    \label{Leff}
\end{equation}
which, in general, is different from $a^*$ (Fig.~\ref{fig:2}a-b). At stationarity, $I_{xy}$ can be estimated similarly to Eq.~\eqref{info}. Analogously, the dissipation on $x$ and $y$ can be obtained from Eq.~\eqref{sdot}, where $\mathbf{\Sigma}$ and $\mathbf{D}$ have to be substituted by the submatrices $\mathbf{\Sigma}_{xy}$ and $\mathbf{D}_{xy}$, and $\mathbf{A}$ by the reduced interaction matrix,
\begin{equation}
    \mathbf{A}_{xy}^{\rm red} = \mathbf{A}_{xy} + \mathbf{C} \mathbf{\Sigma}^{-1}_{xy} \;, \quad (\mathbf{C})_{ij} = (\mathbf{A})_{i\eta} (\mathbf{\Sigma})_{j\eta} \;,
    \label{redA}
\end{equation}
with $i, j = x, y$. $\mathbf{A}_{xy}^{\rm red}$ is the interaction matrix appearing in the Fokker-Planck equation associated with Eq.~\eqref{model}, marginalized over the unobservable DOF, $\eta$ (see Supplemental Material \cite{supplemental_material}).

To evaluate the performance of transduction, we compare the mutual information between the probe and the bath in the ideal but unrealizable case, $I_{y\eta}(a^*)$, with the same quantity when the coupling takes its effective value, $I_{y\eta}(a^*_{\rm eff})$ (Fig.~\ref{fig:2}c). We name $I_{y\eta}(a^*)$ target information, while $I_{y\eta}(a^*_{\rm eff})$ is the transduced information that the probe can effectively harvest. We find that there exists a range of $\sigma$ for which the transduced information is greater than the target one (Fig.~\ref{fig:2}d). Since $a^*_{\rm eff}$ has been set by optimizing the accessible quantities in $\mathcal{L}_{\rm eff}$, this result is unexpected and shows that transduction mechanisms may boost information over the ideal scenario. However, increasing $\sigma$, the membrane dynamics becomes dominated by the bath, inducing, in principle, a larger probe-membrane dissipation that cannot be counterbalanced by $I_{xy}$. As a consequence, $a^*_{\rm eff}$ decreases to minimize $\dot{S}_{xy}$, and transduction first becomes inefficient and then disappears (Fig.~\ref{fig:2}d).

Yet, information transduction happens at a cost. The variance of $I_{y\eta}$, computed as
\begin{equation}
    {\rm var}(I_{y\eta}) = \frac{\Sigma_{y\eta}^2}{\Sigma_{yy} \Sigma_{\eta\eta}} \;,
\end{equation}
is also higher for efficient transduction, evidencing a reduction in processing precision (Fig.~\ref{fig:2}e) (see Supplemental Material \cite{supplemental_material}). Analogously, the system exhibits higher dissipation in the same region of parameters, since harvesting more information requires more energy than in the ideal case (Fig.~\ref{fig:2}f). Overall, once the probe fixes the balance between dissipation and information by choosing a strategy $\lambda$, the possibility of tuning the energy consumption is crucial for the onset of efficient transduction. Different choices of $\lambda$ will change the value of $\sigma$ at which transduction becomes inefficient, without qualitatively affecting the results. Let us comment that, at small values of $\theta_\eta$, the active bath becomes a fast variable, and thus its mutual information with both $x$ and $y$ becomes negligible \cite{nicoletti2021mutual, nicoletti2023propagation}. As a consequence, $a^*$ is tuned to mainly minimize $\dot{S}_a$, leading to a vanishingly small target information and dissipation in the ideal case. Remarkably, this implies that the transduced information still surpasses the target one for a given range of $\sigma$, regardless of the value of $\theta_\eta$ (Fig.~\ref{fig:2}a-c).

Another pivotal complication that constrains operations of a biological system is that it usually must tune its parameters from stochastic trajectories observed on the fly. Conversely, the functional $\mathcal{L}_{\rm eff}$ solely contains averaged quantities (Eqs.~\eqref{info} and \eqref{sdot}). In realistic scenarios, mutual information must instead be computed from stochastic trajectories as
\begin{equation}
\label{eqn:traj_information}
    i_{xy}(t) = \log\frac{p(x(t), y(t))}{p(x(t))p(y(t))}
\end{equation}
where $p(x(t),y(t))$ is the joint pdf between $x$ and $y$ evaluated on each point of the trajectory $\{x(t), y(t)\}$, $p(x(t))$ and $p(y(t))$ the two marginalized distributions. Clearly, $I_{xy} = \langle i_{xy} \rangle$ and ${\rm var}(I_{xy}) = \langle i^2_{xy} \rangle - \langle i_{xy} \rangle^2$, where the average is performed over stationary trajectories \cite{parrondo2015thermodynamics,dabelow2019irreversibility}. Analogously, the probe can estimate dissipation along a single trajectory as:
\begin{equation}
\label{eqn:traj_dissipation}
    \dot{s}_{xy}(t) = \mathbf{A}_{xy}^{\rm red} \circ \left(\dot{x}(t), \dot{y}(t)\right)^T \;,
\end{equation}
where $\circ$ indicates the Stratonovich product \cite{gardiner}. We immediately have that $\dot{S}_{xy} = \langle \dot{s}_{xy} \rangle$ by plugging into Eq.~\eqref{eqn:traj_dissipation} the reduced interaction matrix between $x$ and $y$. Therefore, we can define an effective functional that depends on the observation time $T_A$:
\begin{equation}
\label{eqn:Ltr}
    \mathcal{L}_{\rm tr}(a, T_A) = \lambda \langle i_{xy} \rangle_{T_A} + (1-\lambda) \langle \dot{s}_{xy} \rangle_{T_A}
\end{equation}
where averages are performed over trajectories of duration $T_A$. We name $a_{\rm tr}$ the coupling that maximizes $\mathcal{L}_{\rm tr}$.

\begin{figure}
    \centering
    \includegraphics[width=\columnwidth]{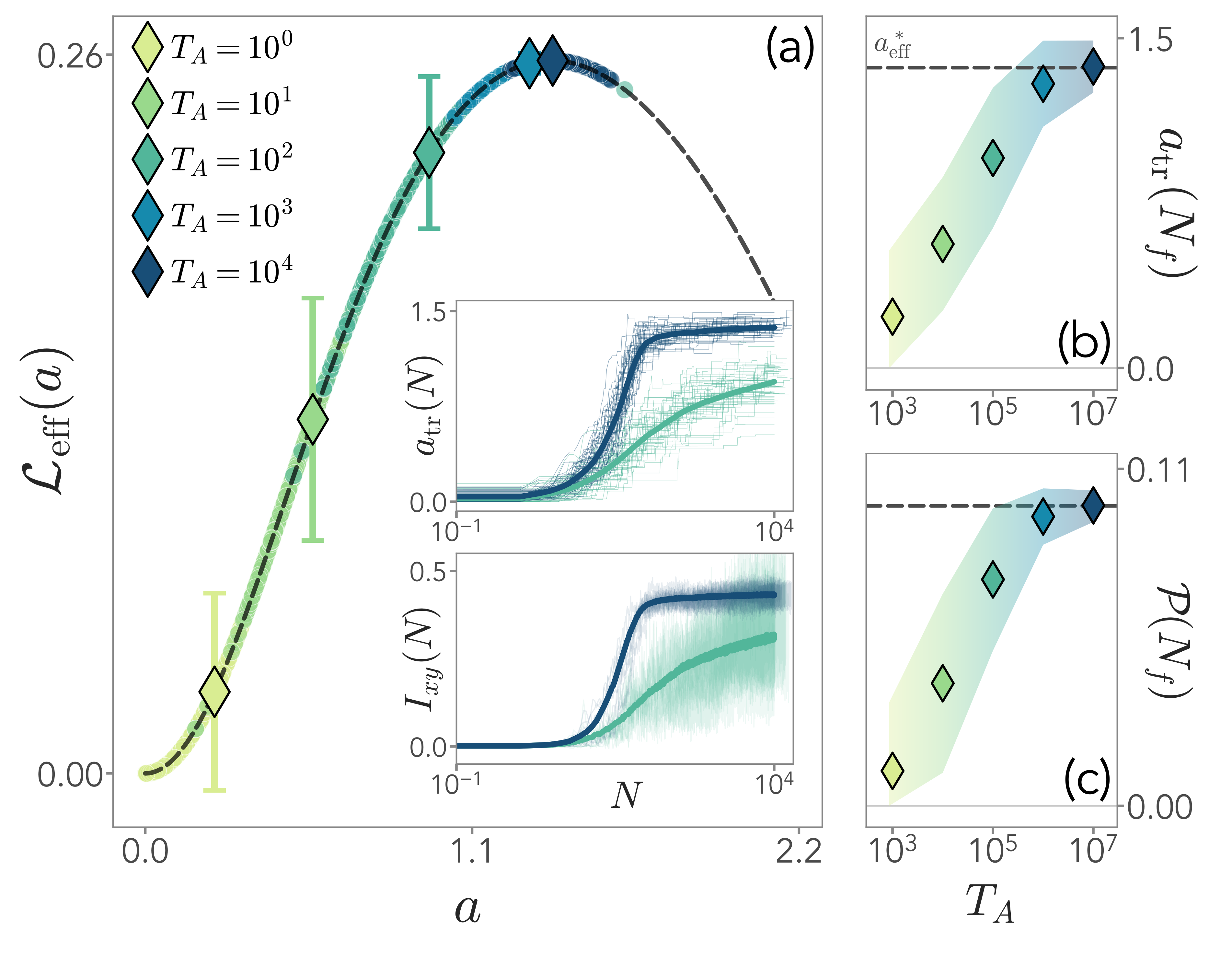}
    \caption{(a) By increasing the observation time, $T_A$, the adaptive dynamics of the probe approaches the maximum of the Pareto functional $\mathcal{L}_{\rm tr}$, Eq.~\eqref{eqn:Ltr}. Diamonds represents average values of the functional, and insets show the optimal coupling and estimated information for two values of $T_A$ versus the number of adaptive steps $N$ for $320$ realizations. (b-c) Optimal coupling (b) and information precision (c) estimated from stochastic trajectories at the final step $N = N_f$ for increasing $T_A$. The shaded area indicates variance.}
    \label{fig:3}
\end{figure}

Consider now a biological probe that continuously adapts the probe-membrane coupling, $a_{\rm tr}(N)$, where the argument $N$ indicates the number of adaptive steps it has undergone. The probe stochastically selects a new coupling $a_{\rm tr}(N) + \Delta a$, with $\Delta a \sim \mathcal{N}(0,\sigma_a)$. During a time $T_A$, it measures information and dissipation via Eq.~\eqref{eqn:traj_information} and Eq.~\eqref{eqn:traj_dissipation}, respectively, resulting in an estimate of $\mathcal{L}_{\rm tr}(a, T_A)$. Based on these temporally limited observations, the adaptive probe tests the performance of its coupling: if the estimated $\mathcal{L}_{\rm tr}(a, T_A)$ is larger than the previous one, it retains the proposed coupling, i.e., $a_{\rm tr}(N+1) =  a_{\rm tr}(N) + \Delta a$; otherwise it discards it, and $a_{\rm tr}(N+1) = a_{\rm tr}(N)$. The dynamics proceeds until no new adaptive move is made for a sufficiently long stopping time, and we assume that the system begins with no probe-membrane coupling. The implementation of this adaptive dynamics (whose details are given in the Supplemental Material \cite{supplemental_material}) leads the system to converge to a coupling that approaches the maximum of $\mathcal{L}_{\rm eff}$ with increasing measurement times $T_A$ (Fig.~\ref{fig:3}a-b). It is worth noting that, as $T_A$ increases, both the number of adaptive steps until convergence and the variance among different realizations decrease (insets of Fig.~\ref{fig:3}a). Moreover, the precision of information transduction,
\begin{equation}
    \mathcal{P}(N) = \frac{\langle i_{y\eta} \rangle^2_{T_A}}{\langle i^2_{y\eta} \rangle_{T_A} - \langle i_{y\eta} \rangle^2_{T_A}} \;,
\end{equation}
increases as well (Fig.~\ref{fig:3}c). Thus, this trajectory-dependent approach allows us to elucidate the basic principles of how biological systems tune their internal parameters directly from observable stochastic trajectories of finite duration.

\begin{figure}
    \centering
    \includegraphics[width=\columnwidth]{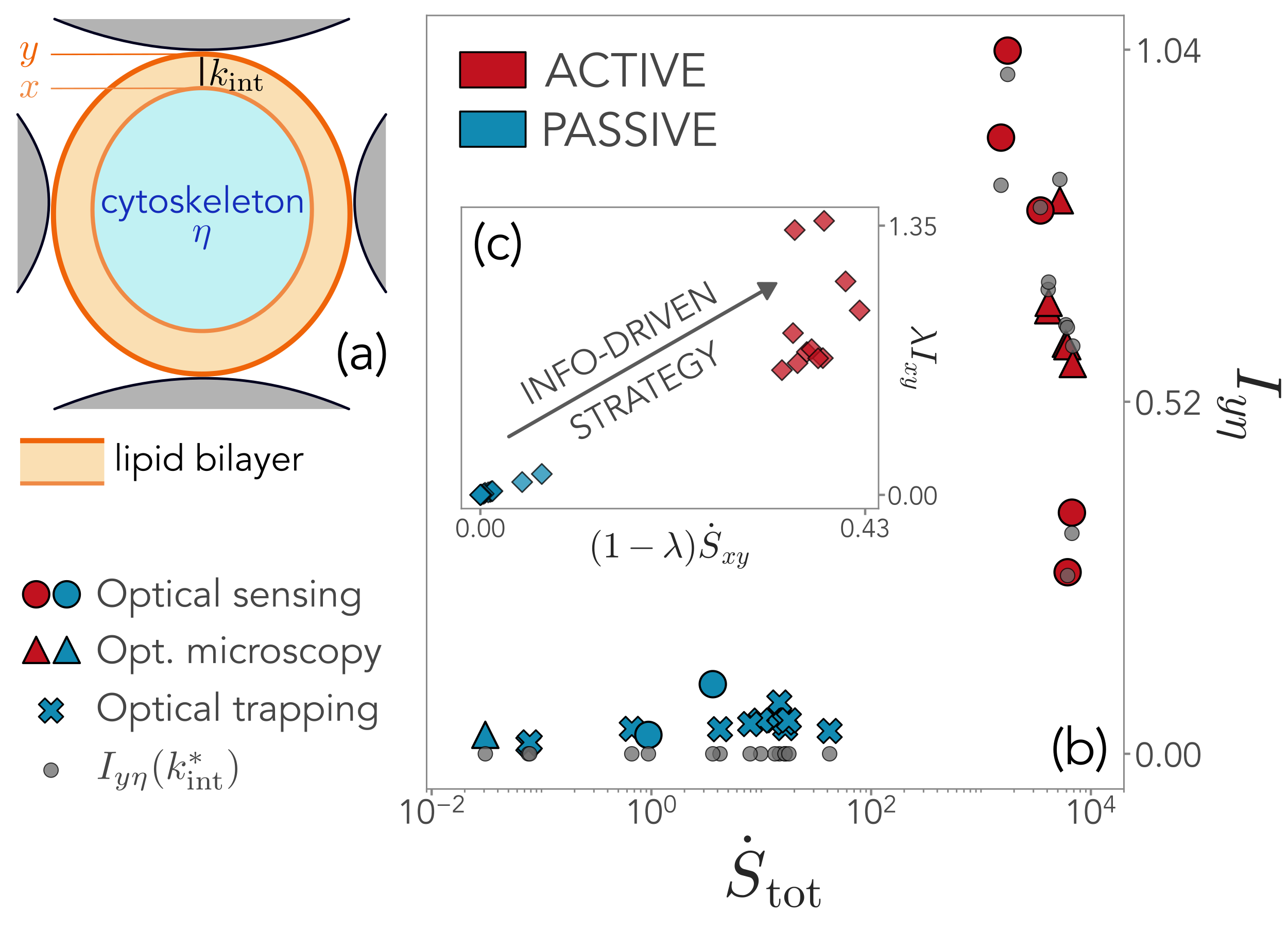}
    \caption{(a) Sketch of the model for RBCs. Cytoskeleton activity, $\eta$, is transduced into the flickering of the outer membrane, $y$, through the inner membrane, $x$, of the lipid bilayer. (b) Information-dissipation balance for passive and active RBCs in different experiments. (c) Increasing dissipation, information becomes more important for the transduction strategies.}
    \label{fig:4}
\end{figure}

Finally, we employ our framework to study transduction in red blood cells (RBCs). Membrane flickering of RBCs is a well-known phenomenon \cite{browicz1890further} that takes place out of equilibrium in healthy cells and, as such, dissipates energy \cite{turlier2016equilibrium}. Various studies reported that flickering might tune cell-cell interactions \cite{evans1986thermal}, favor protein mobility on the membrane \cite{brown2003regulation,lin2004dynamics}, and protein uptake \cite{ayala2023thermal}. All these processes are instrumental in establishing robust cellular functions, and their mechanisms are encoded in the motion of the outer membrane. Therefore, a pivotal question is to understand when and how membrane flickering carries information on the internal active cytoskeleton, transduced through the lipid bilayer. Following a recent model that has been shown to quantitatively capture dynamic and thermodynamic properties of RBCs flickering \cite{di2023variance}, we modify Eq.~\eqref{model} to account for a reciprocal interaction $k_{\rm int}$ between the inner ($x$) and outer ($y$) membrane layer (Fig.~\ref{fig:4}a). We also include the mobility of both membrane layers, $\mu_x$ and $\mu_y$, and we enforce the fluctuation-dissipation relation to fix the diffusion coefficients, $D_x = \beta^{-1} \mu_x$ and $D_y = \beta^{-1} \mu_y$, where $\beta$ is the Boltzmann factor. Cytoskeleton is again modeled as an active noise with a typical timescale, as in \cite{di2023variance}, and its dynamics is known to affect membrane flickering \cite{rodriguez2015direct}. The resulting set of equations is shown in the Supplemental Material \cite{supplemental_material}.

Experimental data from optical sensing, microscopy, and trapping allow for direct measurements of the coupling $k_{\rm int}$ between the membrane layers constituting the lipid bilayer. Using such measures, we first recover the results of \cite{di2023variance} and find a net distinction between the heat dissipated by passive and healthy (active) RBCs. Crucially, we also find that active RBCs exhibit substantially higher values of the mutual information $I_{y\eta}$ between the outer membrane, $y$, and the cytoskeleton, $\eta$, indicating that information is being transduced through the lipid bilayer (red points in Fig.~\ref{fig:4}a).

According to our framework, transduction stems from an underlying strategy $\lambda$ that quantifies how much each RBC weighs information over dissipation, shaping the effective functional in Eq.~\eqref{Leff} to be maximized. Therefore, we can solve the inverse problem of determining the strategy $\lambda$ for each RBC such that their measured coupling $k_{\rm int}$ corresponds to the maximum of $\mathcal{L}_{\rm eff}$. The first result is that the strategy becomes more information-driven, i.e., $\lambda I_{xy} \gg (1-\lambda) \dot{S}_{xy}$, as we increase dissipation going towards active RBCs (Fig.~\ref{fig:4}c). This shows that the internal parameters are tuned so that, the higher the available energy, the greater the information that membrane flickering harvests from the active cytoskeleton. As a second result, we find an intriguing relationship between mechanical stress and transduction efficiency. First, we compute the target information, $I_{y\eta}(k_{\rm int}^*)$, at the same value of $\lambda$ (gray dots in Fig.~\ref{fig:4}b). Here, $k_{\rm int}^*$ is the optimal coupling that the RBC would choose if the outer membrane had direct access to the state of the cytoskeleton, i.e., if it maximized Eq.~\eqref{eqn:functional}. This comparison is analogous to the one shown in Fig.~\ref{fig:2}d. Among active RBCs, we find that transduction is inefficient in experiments performed with optical sensing, while optical microscopy is associated with an efficient regime (Fig.~\ref{fig:4}b). Indeed, in these two settings, cells are held in their position with different setups (details about the experiments can be found in \cite{di2023variance}), hence exhibiting different mechanical properties that are known to crucially affect their functional behavior \cite{dao2003mechanics,suresh2006mechanical,tomaiuolo2014biomechanical}. This naturally translates into a change in the transduction performance, which is immediately highlighted by our framework, suggesting a promising path for further functional investigations of biological systems in different conditions.

In this Letter, we uncovered how biological systems can harvest information on hidden DOFs by tuning their couplings to accessible observables. We found that, in certain regimes, transduced information can overcome the ideal case in which all the DOFs are known, even when only finite-time stochastic trajectories can be measured. These results hint at the existence of intrinsic and crucial constraints on internal parameters of biological systems that allow for efficient transduction mechanisms.
We applied our framework to evaluate, for the first time, how membrane flickering in red blood cells transduces information from the cytoskeleton to the outer membrane, revealing an intriguing link between mechanical stress and transduction performances.

The presented theory is prone to extensions to widely different biological and living organisms. Most importantly, it allows for a deeper characterization of the relation between systems' activity and functional behaviors. Indeed, although dissipation is usually associated with healthy conditions, it is not necessarily tangled with robust cellular function, such as efficient transduction of information. Ultimately, our work sheds light on how non-equilibrium conditions and information harvesting shape adaptive strategies in biological systems.

\begin{acknowledgments}
G.N. acknowledges funding provided by the Swiss National Science Foundation through its Grant CRSII5\_186422. The authors thank Matteo Ciarchi for useful comments and Ivan Di Terlizzi for insightful discussions on the RBC model. They also thank the Max Planck Institute for the Physics of Complex Systems for hosting G.N. during the final phase of this work.
\end{acknowledgments}


\end{document}